\documentclass{article}

\usepackage{PRIMEarxiv}

\usepackage[utf8]{inputenc} 
\usepackage[T1]{fontenc}    
\usepackage{hyperref}       
\usepackage{url}            
\usepackage{booktabs}       
\usepackage{amsfonts}       
\usepackage{nicefrac}       
\usepackage{microtype}      
\usepackage{lipsum}
\usepackage{fancyhdr}       
\usepackage{graphicx}       
\usepackage{amsmath}
\usepackage{threeparttable}
\usepackage{float}
\usepackage{booktabs}
\usepackage{graphicx}
\usepackage{booktabs}
\usepackage{multirow}
\graphicspath{{media/}}     

\title{SPR:Supervised Personalized Ranking Based on Prior Knowledge for Recommendation
}

\author{
  Chun Yang \\
  School of Automation Engineering\\ University of Electronic Science and Technology of China \\
  Chengdu, Sichuan, China\\
  \texttt{beiluo@std.uestc.edu.cn} \\
  \And
  Shicai Fan \\
  School of Automation Engineering\\ University of Electronic Science and Technology of China \\
  Chengdu, Sichuan, China\\
  \texttt{shicaifan@uestc.edu.cn} \\
}

\begin{document}
\maketitle

\begin{abstract}
The goal of a recommendation system is to model the relevance between each user and each item through the user-item interaction history, so that maximize the positive samples score and minimize negative samples. Currently, two popular loss functions are widely used to optimize recommender systems: the pointwise and the pairwise. Although these loss functions are widely used, however, there are two problems. (1) These traditional loss functions do not fit the goals of recommendation systems adequately and utilize prior knowledge information sufficiently. (2) The slow convergence speed of these traditional loss functions makes the practical application of various recommendation models difficult.

To address these issues, we propose a novel loss function named Supervised Personalized Ranking (SPR) Based on Prior Knowledge. The proposed method improves the BPR loss by exploiting the prior knowledge on the interaction history of each user or item in the raw data. Unlike BPR, instead of constructing <user, positive item, negative item> triples, the proposed SPR constructs <user, similar user, positive item, negative item> quadruples. Although SPR is very simple, it is very effective. Extensive experiments show that our proposed SPR not only achieves better recommendation performance, but also significantly accelerates the convergence speed, resulting in a significant reduction in the required training time.

\end{abstract}


\section{INTRODUCTION}
With the development of internet technology, the problem of information overload has become increasingly serious. Recommender systems emerge as a kind of information filter to present content that may be of interest to users\cite{covington2016deep,zhou2019deep,pi2019practice,grbovic2018real}.Recommender systems generally use embedding vectors to represent users and items. The basic idea is that users with similar interaction history have similar interests and preferences, that is, the embeddings of similar nodes should be close to each other in the embedding space. \cite{he2017neural,ebesu2018collaborative,liang2018variational,hernando2016non}.

In order to learn valuable and reliable embedding vectors, a lot of work has been carried out on loss functions in recommender systems \cite{xue2017deep,zhao2021autoloss,mehta2017review}. The loss functions widely used in recommendation systems can be divided into two categories: pointwise loss functions and pairwise loss functions \cite{saito2020unbiased}. Point-wise loss functions such as Mean Squared Error (MSE), Binary Cross-Entropy (BCE) directly fit the label of each sample. Koren et al used MSE to optimize their matrix factorization model \cite{koren2009matrix}. Xue et al. considered both explicit ratings and implicit feedback to improve BCE to optimize their models \cite{xue2017deep}. He et al. used the BCE loss function to optimize their NeuMF model and achieved good results. In addition, there are many models in the industry that use pointwise loss functions to optimize their models \cite{zhou2019deep,zhou2018deep,covington2016deep}. 

Although the pointwise loss function is simple and effective, it can only optimize a single interaction without ranking information, which is not conducive to recommender systems to generate recommendation lists. To address this problem, pairwise loss functions such as BPR \cite{rendle2012bpr} and Collaborative metric learning (CML) \cite{hsieh2017collaborative} have been proposed. Wang et al. optimized their NGCF model using BPR. He et al. used BPR to optimize their lightGCN model with excellent results \cite{he2020lightgcn}. Tay et al. used CML to optimize their LRML model.

\begin{figure*}
	\centering
		\includegraphics[scale=.51]{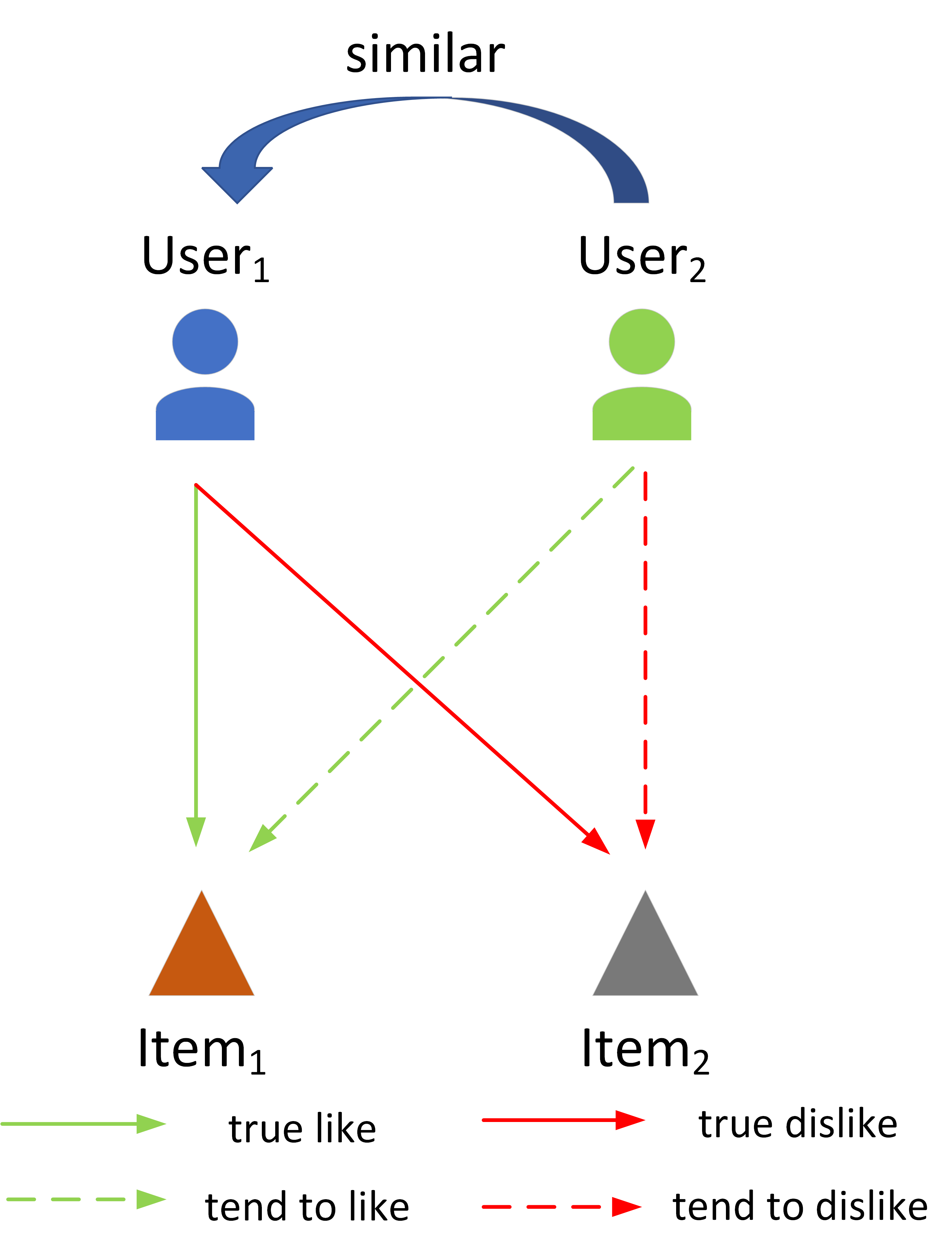}
	\caption{The motivation for SPR. We aim to make nodes with similar interaction history close to each other in the representation space, that is, similar nodes have similar interest tendencies. Specifically, user 1 likes item 1 and dislikes item 2. At the same time, user 2 is similar to user 1, so we think that user 2 also tends to regard item 1 as a positive sample and item 2 as a negative sample.}
	\label{toy_spr}
\end{figure*}

Although these loss functions have achieved widespread success, they suffer from two drawbacks:
\begin{itemize}
\item The above loss functions do not fully utilize prior knowledge of the raw data. Specifically, traditional loss functions directly model the observed implicit interaction data while ignoring the prior knowledge in the interaction data. Using this prior knowledge, we can purposefully make users or items with similar interaction histories learn similar representations to improve recommendation performance. Furthermore, traditional loss functions that directly model the observed implicit interactions are actually modeling the actual click-through rate rather than the correlation between users and items, because having interactions between users and items means that The user observes the item while being interested in the item. When there is no interaction between the user and the item, it does not mean that the user is not interested in the item.
\item The above loss functions do not fully utilize prior knowledge, resulting in their low training efficiency, which makes the practical application of recommendation models very difficult. Specifically, taking LightGCN as an example, it needs 1000 epochs as the optimal parameter to converge. When training with GTX3090, the training process takes about 36 hours.
\end{itemize}

To address the above issues, we propose a Supervised Personalized Ranking (SPR) loss based on prior knowledge. Its design motivation is in line with our previous work (Supervised Contrastive Learning for Recommendation) \cite{yang2022supervised}. That is, nodes with similar interaction histories should have similar representations, which is in line with the motivation of recommender systems. Our previous work has demonstrated that leveraging this prior knowledge can effectively improve representation quality to improve recommendation performance.

Specifically, as shown in Fig. 1, the design motivation of SPR is to use the prior knowledge of the raw data to consider that users with similar interaction history also have the same interest tendency, so that the representations of similar nodes can be close to each other. Even if there is no interaction between user b and item i, we still regard it as a positive sample pair and regard user b and item j as a negative sample pair. Instead of modeling interactions with click history directly, we believe that this modeling approach can model user interest tendencies rather than click-through rates. In order to realize SPR, we adopt the method of dynamic sampling to improve the sampling efficiency.

In addition, compared with the traditional loss function, because that SPR utilizes more prior knowledge, the convergence speed of SPR is significantly faster. This makes the practical application of some complex recommendation models such as graph neural networks less difficult. 

It is worth noting that despite the motivations of SPR is simple, it is very effective. SPR can be easily and directly applied to the backbone networks of various recommender systems proposed in the past to replace loss functions such as BPR.

To summarize, The main contributions of this paper are summarized as follows:
\begin{itemize}
\item We propose a novel loss function named supervised personalized ranking loss based on prior knowledge, which utilizes the prior knowledge of the raw data to model the user's interest tendency, improve the quality of the learned representation, and improve the recommendation performance.
\item Since the proposed SPR utilizes more prior knowledge and relies on the dynamic sampling method, its convergence speed is greatly accelerated, which reduces the difficulty of practical application of complex models.
\item We conduct fully contrastive experiments on various widely used loss functions and backbone networks. The experimental results demonstrate the superior performance of the proposed SPR loss.
\end{itemize}

The rest of this paper is organized as follows. Section \ref{PRELIMINARIES} Details of pairwise and pairwise loss functions. In Section \ref{method}, We detail the details and implementation of SPR. In Section \ref{experiment}, extensive experiments with various loss functions and backbone networks demonstrate the effectiveness and efficiency of the proposed SPR. Finally, conclusions are drawn in Section \ref{conclu}.

\section{PRELIMINARIES }
\label{PRELIMINARIES}

\subsection{Problem Statement}
\label{PROBLEM}

The set of user-item interactions can be easily modeled as a bipartite graph $ {\cal G} = ({\cal U},{\cal V},{\cal E})\ $, where $ {\cal U} $ represents the user set, $ {\cal V} $ represents the item set, and $ {\cal E} $ represents the edge set. If there is an interaction between a user and a item, then there is an edge between them \cite{sun2020neighbor}. The recommendation system needs to embed each user $ i \in {\cal U} $ and item $ j \in {\cal V} $ into a unified d-dimensional representation space, and the corresponding representations are expressed by $ u \in {^d} $ and $ v \in {^d} $. Using the embedded representation, the recommender system can determine the user's interest in a certain item through the score function, so as to decide whether the item should be recommended to the user. The purpose of the recommender system is to make the score of positive sample pairs as large as possible and the score of negative sample pairs as small as possible.

\subsection{Pointwise Loss}
\label{Pointwise}

The pointwise loss function optimizes the observed interaction directly, i.e. optimizes the observed user-item interactions and ground-truth labels. Typical pointwise loss functions are MSE loss function and BCE loss function. The form of the MSE loss function is as follows \cite{chen2020efficient,zhang2018metric}:
\begin{equation}
\begin{aligned}
& {{\cal L}_{MSE}} = \frac{1}{M}{\sum\limits_{u = 1}^M {\left( {{{\hat y}_{ui}} - {y_{ui}}} \right)} ^2} + \lambda {L_2}
\label{mse}
\end{aligned}
\end{equation}

Where $ M $ is the total number of interactions, $ {\hat y}_{ui} $ is the predicted score of the recommendation model for user u and item i, and $ y_{ui} $ is the actual score of user u and item i. The $ \lambda $ is the parameter that controls the intensity of L2 regularization. The MSE loss function directly models the error of predict results and labels, and optimizes the recommended model parameters by minimizing this error.

Unlike the MSE loss, which regards the optimization target as a regression task, the BCE loss regards the optimization target as a classification task, and its form is as follows \cite{he2017neural} :
\begin{equation}
\begin{aligned}
& {{\cal L}_{BCE}} =  - \sum\limits_{u = 1}^M {{y_{ui}}} \ln \sigma ({\hat y}_{ui}) + (1 - {y_{ui}})\ln (1 - \sigma ({\hat y}_{ui})) + \lambda {L_2}
\label{bce}
\end{aligned}
\end{equation}

Although the pointwise loss function can achieve good results, its optimization process does not contain ranking information, which is not conducive to the recommendation model to generate the recommendation list. In general, pairwise loss functions can achieve better results than pointwise loss functions.

\subsection{Pairwise Loss}
\label{pairwise}

The pairwise loss function generally selects a <user, positive item, negative item> to form a triplet, where the positive item represents that the item has interacted with the user, and the negative item represents that the item has no interaction with the user. The optimization goal of the pairwise loss function is to make score of users and positive items higher than that of users and negative items. Typical pairwise loss functions include BPR, CML. The loss function of BPR is as follows \cite{rendle2012bpr}:
\begin{equation}
\begin{aligned}
& {{\cal L}_{BPR}} =  - \sum\limits_{u = 1}^M {\sum\limits_{i \in {N_u}} {\sum\limits_{j \notin {N_u}} {\ln } } } \sigma \left( {{{\hat y}_{ui}} - {{\hat y}_{uj}}} \right) + \lambda {L_2}
\label{bpr}
\end{aligned}
\end{equation}

Where $ {{N_u}} $ represents the set of positive items for user $ i $. BPR can make positive pairs score larger than negative pairs, resulting in better performance of the recommendation model.

Unlike BPR, which directly optimizes the score of user-item pairs, the optimization goal of CML is the distance between users and items in the representation space, and its form is as follows:

\begin{equation}
\begin{aligned}
& {{\cal L}_{CML}} = \sum\limits_{u = 1}^M {\sum\limits_{i \in {N_u}} {\sum\limits_{j \notin {N_u}} {{{[m + d{{(u,i)}^2} - d{{(u,j)}^2}]}_ + }} } } + \lambda {L_2}
\label{cml}
\end{aligned}
\end{equation}

where $ {d{{(a,b)}^2}} $ denotes the euclidean distance between node $ a $ and node $ b $, and $ {[z]_ + } = max(z,0) $ denotes the standard hinge loss. In addition, CML loss also includes feature loss, details about CML can be viewed in \cite{hsieh2017collaborative}. 

Although the performance of the pairwise loss function has been excellent, we believe that the above loss does not fully utilize the prior knowledge of the original interaction history data. Furthermore, we believe that the above loss functions cannot adequately express the correlation between users and items. Therefore, we propose the SPR loss function.

\section{METHODOLOGY}
\label{method}

\subsection{The Pipeline Of SPR}
\label{pipeline}

The reason for we propose SPR is that traditional loss functions do not adequately model the user's interest tendencies. Specifically, for a user i and an item j, there are interaction variables $ {Y_{ij}} $, correlation variables $ {R_{ij}} $, and observation variables $ {O_{ij}} $. When there is interaction between the user and the item, $ {Y_{ij}} = 1 $, otherwise 0, when the user is related to the item, $ {R_{ij}} = 1 $, otherwise 0, when the user observes the item, $ {O_{ij}} = 1 $, otherwise 0. Only when the user i is associated with the item Interactive behavior occurs when item j is related and there is an observation behavior. Specifically, the mathematical form is:
\begin{equation}
\begin{aligned}
& P({Y_{ij}} = 1) = P({R_{ij}} = 1)P({O_{ij}} = 1|{R_{ij}} = 1)
\end{aligned}
\end{equation}

This means that the traditional loss function models the interaction variable $ {Y_{ij}} $, not the correlation variable $ {R_{ij}} $. Because in the historical interaction data, we can only directly observe the interaction history, but not the correlation. Furthermore, traditional loss functions do not fully utilize prior knowledge, they only utilize user-item relationships information, not user-user relationships information.

In order to solve the above problems, this paper proposes the SPR loss function. SPR not only utilizes the user-item relationship information, but also fully considers the user-user relationship under the constraint of prior knowledge. Specifically, instead of building <$ user $, $ positive item $, $ negative item $> triples, SPR builds <$ use{r_i} $, $ use{r_l}  $, $ positive item $, $ negative item $> quadruple, where $ use{r_l}  $ has similar interaction histories with $ use{r_i} $. SPR follows the basic idea of recommendation systems, that is, users with similar interaction histories have the same interest tendency. Therefore, it is considered that $ use{r_l}  $ has the same interest tendency as $ use{r_i} $, and the positive sample of $ use{r_i} $ is regarded as the positive sample of $ use{r_l}  $, even if there is no real interaction between them. Similarly, the negative sample of $ use{r_i} $ is regarded as the negative sample of $ use{r_l}  $.

In order to introduce the prior knowledge of the original interaction data, SPR first needs to calculate the similarity between each user, The matrix form of similarity calculation is as follows:
\begin{equation}
\begin{aligned}
& S = \frac{{{\cal G}{{\cal G}^T}}}{{\left| {\cal G} \right|}}
\label{s}
\end{aligned}
\end{equation}

Where $ {\cal G} $ is the sparse matrix which represents the interaction history of each user. $ {\left| {\cal G} \right|} $ means row normalization of the matrix. After computing the similarity, we sort each row of $ S $ and pick the top $ k\% $ most similar users. 

Then for each node we will sort $ S $ and select the top $ k\% $ nodes as the similar nodes that can be used for replacement. It is worth noting that when calculating similarity, user-side nodes and item-side nodes are calculated separately. Assuming that there are $ N $ users in the entire interaction history, then for each $ use{r_t} $, we will select $ N*k\%  $ users as the similar user of the $ use{r_t} $.

Then, the SPR loss can be calculated as:
\begin{equation}
\begin{aligned}
& {{\cal L}_{SPR}} =  - \sum\limits_{u,l,i,j \in D} {\ln \sigma ({{\hat y}_{ui}} - {{\hat y}_{uj}} + {{\hat y}_{li}} - {{\hat y}_{lj}})}
\end{aligned}
\end{equation}

Where $ {Y_{ui}} = 1 $ ,$ {Y_{uj}} = 1 $ and $ l $ is a random similar user of $ i $ obtained by equation \ref{s}. $ l $ does not require specific interaction with item $ i $ and item $ j $. $ D $ represents the training set of SPR.

It can be observed that we not only consider the user-item relation side but also the user-user relation side in the loss function. The information on the user-user relationship side comes from the prior knowledge of historical interaction data, which enables the model to use more information for learning and accelerates the convergence speed of the model. Furthermore, instead of directly modeling observed interactions, SPR considers user-item pairs without interactions as positive samples, which makes SPR-learned embeddings more likely to represent users' interests rather than click-through rates.

\subsection{Sampling For SPR}
\label{sampling}

Since SPR needs to use both user $ i $ and similar user $ l $ to calculate the loss, how to sample data is crucial for the application of SPR. An intuitive approach is to group all similar users of user $ i $ into user pairs and compute the loss collectively. However, this will bring huge computational cost on large-scale datasets, which is not in line with the purpose of SPR to improve the model convergence speed \cite{rendle2014improving}. In order to ensure the training efficiency of SPR, we follow the basic idea of dynamic sampling for BPR and implement dynamic sampling on SPR \cite{he2020lightgcn,zhang2013optimizing,rendle2012bpr}. The specific steps for sampling for SPR are as follows.

\begin{enumerate}
\itemsep=0pt
\item Determine the batch size $ B $ and the sampling rate $ \gamma $ for model training.

\item For each batch, a list of length $ B*\gamma^{2}  $ is randomly generated, and the content in the list is the user id. The purpose of $ \gamma^{2} $ is to have enough margin to ensure successful sampling.

\item For each user $ u $in the list, randomly sample one positive item $ i $and one negative item $ j $ from the interaction history.

\item Randomly select a user $ l $ from the list of similar users of the current user $ i $, form a <user$ u $, user$ l $, item$ i $, item$ j $> quadruplet, and add the quadruple to the sampling list.

\item If the length of the sampling list is equal to $ B*\gamma $, then output the sampling list and end sampling, otherwise, repeat steps 3-4.
\end{enumerate}

\subsection{Complexity Analyses of SPR}
\label{Complexity}

Compared with BPR, SPR will increase the computational complexity from three aspects. First, calculating the similarity of user interaction history. Assuming that there are $ N $ users in the dataset, the complexity of matrix multiplication is $ O(N^{3})  $, and the complexity of sorting is $ O(\log_{2}{N^{2}}) $. When dynamic sampling is performed, the complexities of SPR and BPR are in same order of magnitude, that is, $ O(N) $. The computational complexity required for BPR sampling is assumed to be $ O(\vartheta)  $, then the computational complexity required for SPR sampling is $ o(\vartheta*\gamma) $. When calculating the loss function, SPR is also in the same order of magnitude as BPR. Assuming that the computational complexity of BPR when calculating the loss is $ O(\tau) $, then the computational complexity of SPR is $ o(2*\tau) $. In general, the additional computation of SPR can be expressed as:
\begin{equation}
\begin{aligned}
& {\Delta} = O({N^{3}})+O({\log_{2}{N^{2}}})+O((\gamma-1)*\vartheta)+O(\tau)
\end{aligned}
\end{equation}

It is worth noting that the calculation of the similarity matrix is carried out in the data preprocessing stage, which means that the computational cost of SPR in the model part is an order of magnitude with that of BPR.

\section{EXPERIMENT}
\label{experiment}

We first illustrate the experimental settings include datasets, evaluation metrics and implementation details in Section \ref{expe_set}. Next, we conduct extensive comparative experiments on a variety of different loss functions and backbone networks in Section \ref{performance}. Finally, we conduct further research on the application of SPR in Section \ref{study}.

\subsection{Experimental Settings}
\label{expe_set}

\subsubsection{Datasets}
\label{data}

Three public recommendation datasets including Gowalla \cite{liang2016modeling,he2016vbpr}, Yelp2018 \cite{wang2019neural,he2020lightgcn} and Amazon-Book \cite{he2016ups} are adopted for top-K recommendation to verify the effectiveness of the SPR. In order to ensure the consistency and fairness of different conditions in the experiment, the training set and test set used in different conditions are exactly the same as those used in the LightGCN paper \cite{he2020lightgcn}. The specific statistics of the three datasets are shown in the Table \ref{datasets}.

\begin{table}
  \centering
  \caption{Statistics of the used three datasets}
    \begin{tabular}{c|c|c|c|c}
    \toprule
    \toprule
    \textbf{Dataset} & \textbf{Users\#} & \textbf{Items\#} & \textbf{Interactions\#} & \textbf{Density} \\
    \midrule
    \midrule
    \textbf{Gowalla} & 29858 & 40981 & 1027370 & 0.00084\% \\
    \midrule
    \midrule
    \textbf{Yelp2018} & 31668 & 38048 & 1561406 & 0.00130\% \\
    \midrule
    \midrule
    \textbf{Amazon-Book} & 52643 & 91599 & 2984108 & 0.00062\% \\
    \bottomrule
    \bottomrule
    \end{tabular}
  \label{datasets}%
\end{table}%

\subsubsection{Experimental Metrics}
\label{metrics}

In top-k recommendation, we implement the strategy as that described in \cite{wang2019neural} and the ranking protocol described in \cite{wang2019neural}. Normalized Discounted Cumulative Gain (NDCG) and Recall as recommended performance indicators are adopted to evaluate the recommendation performance. we consider the case where K to be 20 as LightGCN paper used \cite{he2020lightgcn}.

\subsubsection{Implementation Details}
\label{details}

Here, we compare six loss functions on three different backbone networks. The six loss functions are MSE \cite{koren2009matrix}, BCE \cite{he2017neural,zhou2019deep,covington2016deep}, BPR \cite{he2016vbpr,rendle2012bpr,he2020lightgcn}, CML \cite{hsieh2017collaborative}, User Interest Boundary (UIB) \cite{zhuo2022learning}, SPR (ours). The three backbone networks are MF \cite{koren2009matrix}, NeuMF \cite{he2017neural}, and LightGCN \cite{he2020lightgcn}.

Specifically, the loss functions of MSE, BCE, BPR, and CML are obtained by formula \ref{mse}, \ref{bce}, \ref{bpr} and \ref{cml} respectively. The loss function of UIB has the following form:
\begin{equation}
\begin{aligned}
& {{\cal L}_{UIB}} = {\sum\limits_{i=1}^M}({\sum\limits_{i \in {N_u}}\phi{\left( {b_u - {\hat y}_{ui}} \right)}}+ {\sum\limits_{j \notin  {N_u}}\phi{\left( {{\hat y}_{uj}} - b_u \right)}})+ \lambda {L_2}
\label{uib}
\end{aligned}
\end{equation}

Where $ \phi $ means $ ln\sigma $ here and $ b_u $ represents the personalized decision boundary for user$ i $, which can be calculated as follows:
\begin{equation}
\begin{aligned}
& b_u = W^T*u_i
\end{aligned}
\end{equation}

Where $ W^T $ is the learnable parameters shared by each user, and $ u_i $ is the embedding vector output by the recommendation model for user$ i $. UIB combines the flexibility of pointwise loss function with the advantages of the pairwise loss function for ranking, and can learn personalized decision boundaries for each user, and achieve good results.

MF is the basic collaborative filtering model, which represents each user and item with an embedding vector. It optimizes the embedding representations through matrix factorization. MF directly calculates the correlation between users and items through the inner product of the embedding vectors, and generates a recommendation list.

NeuMF is a collaborative filtering model based deep learning. It not only retains the part of optimizing embedding by matrix factorization, but also proposes the part of network to optimize embedding by MLP. NeuMF takes a user and an item as input and outputs prediction scores directly through a linear layer.

LightGCN is a collaborative filtering algorithm based on graph neural network. LightGCN considers the domain of the current node, and updates the embedding vector of the current node through information transfer and aggregation. It obtains the high-order field information of the node through multiple iterations. LightGCN uses the inner product of embedding vectors to predict user and item scores.

It is worth noting that the CML loss function requires embeddings to calculate the euclidean distance between each node. Applying CML on NeuMF, we take the input of the last linear layer as the embedding representation of the node. In addition, CML optimizes the euclidean distance of the representation, so its prediction score is also determined by the euclidean distance between the user and the item. When the embedding distance is farther, its score is lower.

Specifically, the score functions under different conditions are shown in Table \ref{condition_score}.

\begin{table}[htbp]
  \centering
  \caption{Score functions of different backbone networks under different loss functions.}
    \begin{tabular}{c|c|c}
    \toprule
    \toprule
    \textbf{backbone} & \textbf{Loss} & \textbf{Score} \\
    \midrule
    \multirow{6}[4]{*}{\textbf{MF}} & \textbf{MSE} & \multirow{5}[2]{*}{$ u^Tv $} \\
          & \textbf{BCE} &  \\
          & \textbf{BPR} &  \\
          & \textbf{UIB} &  \\
          & \textbf{SPR(ours)} &  \\
\cmidrule{2-3}          & \textbf{CML} & $ \left | u-v \right |_{2}^{2} $ \\
    \midrule
    \multirow{6}[4]{*}{\textbf{NeuMF}} & \textbf{MSE} & \multirow{5}[2]{*}{$ f(u,v) $} \\
          & \textbf{BCE} &  \\
          & \textbf{BPR} &  \\
          & \textbf{UIB} &  \\
          & \textbf{SPR(ours)} &  \\
\cmidrule{2-3}          & \textbf{CML} & $ \left | f(u)-f(v) \right |_{2}^{2}  $ \\
    \midrule
    \multirow{6}[4]{*}{\textbf{MF}} & \textbf{MSE} & \multirow{5}[2]{*}{$ G(u)^TG(v) $} \\
          & \textbf{BCE} &  \\
          & \textbf{BPR} &  \\
          & \textbf{UIB} &  \\
          & \textbf{SPR(ours)} &  \\
\cmidrule{2-3}          & \textbf{CML} & $ \left | G(u)-G(v) \right |_{2}^{2}  $ \\
    \bottomrule
    \end{tabular}%
  \label{condition_score}%
\end{table}%

To be fair, The embedding size is fixed to 64 for all models and the embedding parameters are initialized with the Xavier method \cite{glorot2010understanding}. The optimizer used for all models is Adam \cite{kingma2014adam}. The batch size is set to 1024, the L2 regularization constant $ \lambda  $ is set to be $ 1 e^{-4} $ and the learning rate is set to 0.001 for all models and loss functions. For parameters unique to SCL, we conducted separate explorations. The $ \gamma $ is searched in the range of $\left\{4, 5, 6,7,8,10,12,15,20\right\}$. $ \gamma $ is set to 10 when compare SPR with other models.

\subsection{Performance Comparison}
\label{performance}

We compare SPR with MSE, BCE, BPR, CML, and UIB with MF, NeuMF, and LightGCN as backbone on Gowalla, Yelp2018 and Amazon-Book dataset, and the results are shown in the Table \ref{performance_compare}. In the table, the best performance is bolded, and the suboptimal performance is underlined. The average improvement represents the relative performance improvement achieved by SPR over the suboptimal results on the three backbone networks.

\begin{table}[htbp]
  \centering
  \caption{Performance comparison with SPR, MSE, BCE, BPR, CML, and UIB with MF, NeuMF, and LightGCN as backbone on Gowalla, Yelp2018 and Amazon-Book dataset. All loss functions and backbone networks are reimplemented according to the corresponding paper to get performance. In the table, the best performance is bolded, and the suboptimal performance is underlined. The average improvement represents the relative performance improvement achieved by SPR over the suboptimal results on the three backbone networks.}
  \resizebox{\textwidth}{!}{
    \begin{tabular}{c|c|c|c|c|c|c|c}
    \toprule
    \toprule
    \multirow{2}[4]{*}{\textbf{Loss function}} & \multirow{2}[4]{*}{\textbf{backbone}} & \multicolumn{2}{c|}{\textbf{Gowalla}} & \multicolumn{2}{c|}{\textbf{Yelp2018}} & \multicolumn{2}{c}{\textbf{Amazon-Book}} \\
\cmidrule{3-8}          &       & \textbf{Recall} & \textbf{NDCG} & \textbf{Recall} & \textbf{NDCG} & \textbf{Recall} & \textbf{NDCG} \\
    \midrule
    MSE   & \multirow{6}[2]{*}{MF} & 0.1502 & 0.1202 & 0.048 & 0.0383 & 0.0308 & 0.0232 \\
    BCE   &       & 0.1516 & 0.1206 & 0.0493 & 0.0385 & 0.0296 & 0.0223 \\
    BPR   &       & 0.1524 & 0.1233 & 0.0489 & 0.0395 & 0.0339 & 0.0258 \\
    CML   &       & 0.1523 & 0.1218 & 0.0485 & 0.0392 & 0.0342 & 0.0259 \\
    UIB   &       & \underline{0.1543} & \underline{0.1238} & \underline{0.0514} & \underline{0.0408} & \underline{0.0346} & \underline{0.0264} \\
    SPR   &       & \textbf{0.1556} & \textbf{0.1262} & \textbf{0.0521} & \textbf{0.0411} & \textbf{0.0377} & \textbf{0.0295} \\
    \midrule
    MSE   & \multirow{6}[2]{*}{NeuMF} & 0.1641 & 0.1301 & 0.0526 & 0.0415 & 0.0337 & 0.0273 \\
    BCE   &       & 0.153 & 0.1236 & 0.0520 & 0.0409 & 0.0348 & 0.0288 \\
    BPR   &       & 0.1672 & 0.1326 & 0.0540 & 0.0434 & \underline{0.0359} & \underline{0.0304} \\
    CML   &       & 0.1675 & 0.1328 & \underline{0.0549} & \underline{0.0436} & 0.0353 & 0.0289 \\
    UIB   &       & \underline{0.1688} & \underline{0.1331} & 0.0547 & 0.0432 & 0.0357 & 0.0295 \\
    SPR   &       & \textbf{0.1698} & \textbf{0.1343} & \textbf{0.0574} & \textbf{0.0460} & \textbf{0.0380} & \textbf{0.0319} \\
    \midrule
    MSE   & \multirow{6}[2]{*}{LightGCN} & 0.167 & 0.1421 & 0.0536 & 0.0435 & 0.0375 & 0.0298 \\
    BCE   &       & 0.1678 & 0.1461 & 0.0538 & 0.0437 & 0.0388 & 0.0302 \\
    BPR   &       & 0.1822 & 0.1544 & 0.0632 & 0.0523 & 0.0410 & 0.0316 \\
    CML   &       & \underline{0.1824} & \underline{0.1548} & 0.0621 & 0.0511 & 0.0428 & 0.0328 \\
    UIB   &       & 0.1822 & 0.1545 & \underline{0.0651} & \underline{0.0530} & \underline{0.0436} & \underline{0.0335} \\
    SPR   &       & \textbf{0.1845} & \textbf{0.1571} & \textbf{0.0674} & \textbf{0.0551} & \textbf{0.0451} & \textbf{0.0355} \\
    \midrule
    \multicolumn{2}{c|}{Average Improve} & 0.86\% & 1.44\% & 3.15\% & 3.40\% & 6.08\% & 7.55\% \\
    \bottomrule
    \end{tabular}}
  \label{performance_compare}%
\end{table}%

From the perspective of different loss function, it can be seen from the table that our proposed SPR which is applied to three backbone networks achieves consistent improvement on different datasets. For Gowalla, Yelp2018 and Amazon-book, the results of recall are improved 0.86\%, 3.15\%, 6.08\% respectively and the results of ndcg are improved 1.44\%, 3.40\%, 7.55\% respectively compared to the suboptimal loss function. Furthermore, since SPR is improved based on BPR, it is necessary to directly compare SPR with BPR. The results of recall are average improved 1.64\%, 6.50\%, 9.02\% respectively and the results of ndcg are improved 1.79\%, 5.13\%, 10.54\% respectively compared to the suboptimal loss function. It can be seen that, compared with BPR, SPR utilizes the prior knowledge of the original data, and more user-user information can be utilized by establishing quadruples, and a very significant improvement has been achieved. It is worth noting that the improvement on the Gowalla dataset is relatively small. We believe this is because Gowalla's smaller data scale makes it harder to introduce diversity when building user-user relationships, which makes information SPR can use less than that on Amazon-book. This shows that SPR is very suitable for application on relatively larger datasets.

From the perspective of the backbone network, it can be seen from the table that our proposed SPR achieves consistent improvement on all backbone networks. Whether it is a traditional matrix decomposition model, a neural network model or a graph convolutional network model, SPR can be applied very simply to improve the effect. This fully demonstrates the flexibility and transferability of SPR. In addition, we can also notice that the performance improvement ratio of SPR is similar on different backbone networks. This means that SPR does not depend on the feature extraction capability of the backbone network, and SRP can stably achieve performance improvement.

In short, we can see from the table \ref{performance_compare}. SPR achieves stable performance improvement on all datasets and all backbone networks. The performance of SPR has a greater relationship with the dataset and less with the backbone network.

\begin{figure*}
	\centering
		\includegraphics[scale=.4]{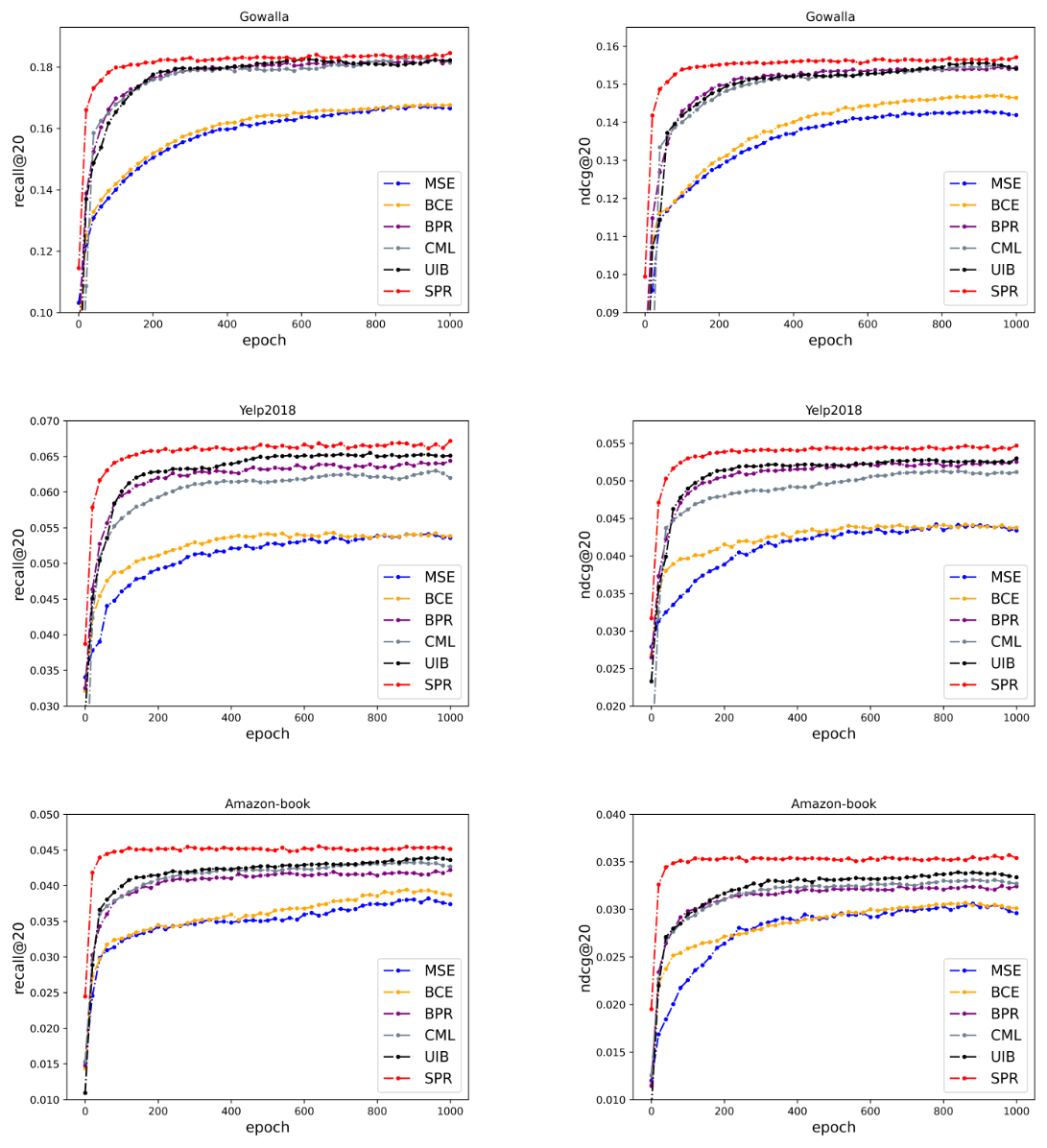}
	\caption{Training curves of different loss functions based on LightGCN, which are evaluated by testing recall and testing ndcg per 20 epochs on Gowalla, Yelp2018 and Amazon-Book.}
	\label{curves}
\end{figure*}

In order to fully illustrate the advantage of SPR in terms of convergence speed, we compare the trends of recall and ndcg with the training process under different loss functions. The train curves which are evaluated by testing recall and testing ndcg per 20 epochs are shown as Figure \ref{curves}. In order to save space, we only show the training curve of each loss function based on LightGCN, and omit the results on MF and NeuMF, which show the same 
trend of that as LightGCN.

As shown in the Figure \ref{curves}, we can observe that the convergence curve of our proposed SPR is quite steep, which means that SPR has a very obvious advantage in the convergence speed. Specifically, BPR, UIB, and CML generally require 1000 epochs to ensure their convergence, which generally takes about 40 hours, 48 hours, and 54 hours on the Gowalla dataset, respectively. While our proposed SPR only needs about 100 epochs to converge to comparable results to BPR, UIB, and CML, which only takes about 5 hours, although SPR requires more sampling. This advantage is even more obvious on the Amazon-book dataset, where SPR only needs 50 epochs to surpass BPR. This is because SPR uses prior knowledge to establish supervised information, and can use more interaction data of the user-item bipartite graph to establish user-user relationships, so that more information is considered in each iteration, making the convergence speed significant accelerate.

In addition, it can also be found from the Figure \ref{curves} that with the further training of SPR, the performance of the recommendation model is further improved. This further illustrates the importance of making full use of prior knowledge. SPR can make the recommendation model converge better and learn higher-quality representations of users and items, thereby improving recommendation performance.

Combining the results in Table \ref{performance_compare} and Figure \ref{curves}, we can draw the following conclusions: (1) First, SPR has better recommendation performance, which outperforms traditional pointwise loss functions and pairwise loss functions as well as some novel loss functions. (2) The convergence speed of SPR is significantly improved, and it only takes a few hours to make the model converge to a satisfactory performance. This makes practical application of some complex models easier. In general, SPR has important application value.

\subsection{Study of SPR}
\label{study}

\begin{figure}
	\centering
		\includegraphics[scale=.45]{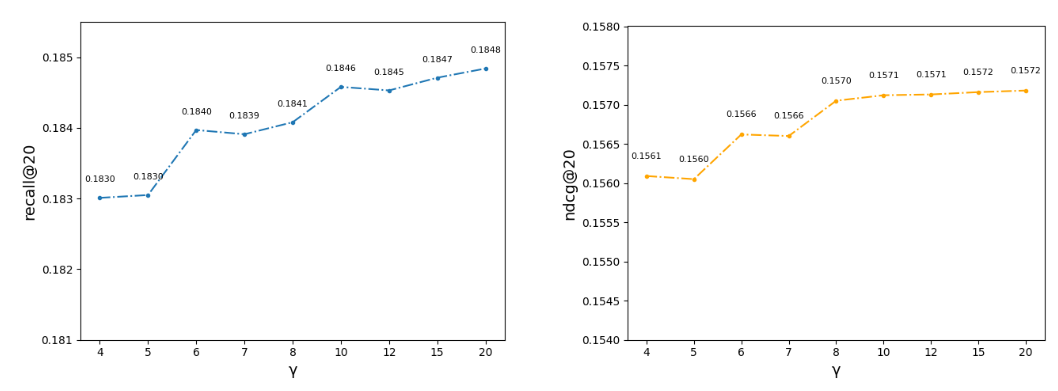}
	\caption{The trend of the performance of SPR as a function of $ \gamma $.}
	\label{effect_gamma}
\end{figure}

The implementation of SPR is very simple. Compared with BPR, its unique parameter is only sampling rate $ \gamma $. To explore the effect of $ \gamma $ on SPR, $ \gamma $ is searched in the range of $\left\{4, 5 ,6,7,8,10,12,15,20\right\}$, and the results are shown in Figure \ref{effect_gamma}. To save space, we omit the experimental results on Yelp2018 and Amazon-book, which show the same trend as that on the Gowalla dataset.

As can be seen from the Figure \ref{effect_gamma}, as $ \gamma $ increases, the performance of SPR will also improve. However, although larger $ \gamma $ will achieve better performance, too large $ \gamma $ will further increase the sampling complexity and loss complexity, which defeats our purpose of improving the convergence speed of the model. Therefore, we set $ \gamma $ to 10 as a compromise and recommend careful tuning of $ \gamma $ when using SPR.

In additon, as mentioned before, SPR is in line with our previous work Supervised Contrastive Learning for Recommendation (SCL). Their basic idea is to make full use of the prior knowledge of the user-item bipartite graph to construct supervised information, and aim to make users or items with similar interaction history learn similar representations. Therefore, the combination of SCL and SPR is a very natural choice. 

We compare SCL-SPR with NGCF \cite{wang2019neural}, LightGCN \cite{he2020lightgcn}, SGL \cite{wu2021self}, SCL \cite{yang2022supervised}. The reason for choosing these models as the baseline is that they are all developed based on NGCF, and it is fair and reasonable to compare them on the same standard. The results are shown in Table \ref{pre_scl_spr}.

\begin{table*}[htbp]
  \centering
  \caption{Performance comparison with NGCF, LightGCN, SGL, SCL and SCL-SPR. Among it, SCL represents the model based on LightGCN. The bold indicates the best result.}
    \begin{tabular}{c|c|c|c|c|c|c}
    \toprule
    \toprule
    \textbf{Database} & \multicolumn{2}{c|}{\textbf{Gowalla}} & \multicolumn{2}{c|}{\textbf{Yelp2018}} & \multicolumn{2}{c}{\textbf{Amazon-Book}} \\
    \midrule
    \textbf{Method} & \textbf{Recall} & \textbf{NDCG} & \textbf{Recall} & \textbf{NDCG} & \textbf{Recall} & \textbf{NDCG} \\
    \midrule
    \midrule
    \textbf{NGCF} & 0.1569 & 0.1327 & 0.0579 & 0.0477 & 0.0337 & 0.0261 \\
    \textbf{LightGCN} & 0.1823 & 0.1555 & 0.0639 & 0.0525 & 0.041 & 0.0318 \\
    \textbf{SGL} & 0.1827 & 0.1558 & 0.0675 & 0.0555 & 0.0478 & 0.0379 \\
    \textbf{SCL} & 0.1833 & 0.1562 & 0.0684 & 0.0565 & 0.0483 & 0.0381 \\
    \textbf{SCL-SPR} & \textbf{0.1853} & \textbf{0.1574} & \textbf{0.0689} & \textbf{0.0567} & \textbf{0.0488} & \textbf{0.0385} \\
    \bottomrule
    \end{tabular}%
  \label{pre_scl_spr}%
\end{table*}%

As shown in the table, combining SCL with SPR resulted in a clear improvement over SCL. Compared with other baselines, the performance advantage of SCL-SPR is obvious. For example, compared with LightGCN, the recall of SCL-SPR on the three datasets has increased by 1.65\%, 7.82\%, 19.24\%, and ndcg has increased by 1.22\%, 8\%, and 21.07\%, respectively. Both SCL and SPR utilize prior knowledge to construct supervised information, resulting in performance improvements. This fully demonstrates the effectiveness and necessity of utilizing prior knowledge.

\section{Conclusion}
\label{conclu}
In this work, we recognize the defect of traditional loss functions that do not fully utilize prior knowledge, and propose a novel loss function called Supervised Personalized Ranking. SPR utilizes the prior knowledge of the interaction history of the user-item bipartite graph to construct supervised information, and constructs <user$ u $, user$ l $, item$ i $, item$ j $> quadruplets for training under the guidance of the supervised information. In addition to considering the user-item relationship, SPR additionally considers the user-user relationship, which enables SPR to achieve better performance. Furthermore, our experiments also demonstrate that SPR has a very significant advantage in convergence speed because SPR utilizes more information. This makes SPR have high application value.

In this work, we use quadruplets to improve recommendation performance. However, SPR is very dependent on sample sampling. How to further improve the loss so that the recommendation loss can consider more samples during training is a very interesting research direction.

\bibliographystyle{unsrt}  
\bibliography{references}

\end{document}